\documentclass[manuscript,screen,review=False,anonymous=False]{acmart}
\AtBeginDocument{%
  }

\usepackage{hhline,colortbl}
\usepackage[dvipsnames]{xcolor}
\usepackage{subcaption}

\setcopyright{none}

\begin{document}

\title[Full Disclosure, Less Trust?]{Full Disclosure, Less Trust? How the Level of Detail about AI Use in News Writing Affects Readers' Trust}

\author{Pooja Prajod}
\orcid{0000-0002-3168-3508}
\email{Pooja.Prajod@cwi.nl}
\affiliation{%
  \institution{Centrum Wiskunde \& Informatica}
  \city{Amsterdam}
  \country{the Netherlands}
}

\author{Hannes Cools}
\affiliation{%
  \institution{University of Amsterdam}
  \city{Amsterdam}
  \country{the Netherlands}}

\author{Thomas Röggla}
\affiliation{%
  \institution{Centrum Wiskunde \& Informatica}
  \city{Amsterdam}
  \country{the Netherlands}
}

\author{Karthikeya Puttur Venkatraj}
\affiliation{%
  \institution{Centrum Wiskunde \& Informatica}
  \city{Amsterdam}
  \country{the Netherlands}
}

\author{Amber Kusters}
\affiliation{%
  \institution{Centrum Wiskunde \& Informatica}
  \city{Amsterdam}
  \country{the Netherlands}
}

\author{Alia ElKattan}
\affiliation{%
  \institution{New York University}
  \city{New York}
  \country{USA}
}

\author{Pablo Cesar}
\affiliation{%
  \institution{Centrum Wiskunde \& Informatica and}
  \institution{TU Delft}
  \country{The Netherlands}
  }

\author{Abdallah El Ali}
\affiliation{%
  \institution{Centrum Wiskunde \& Informatica and}
  \institution{Utrecht University}
  \country{The Netherlands}
  }

\renewcommand{\shortauthors}{Prajod et al.}

\begin{abstract}
As artificial intelligence (AI) is increasingly integrated into news production, calls for transparency about the use of AI have gained considerable traction. Recent studies suggest that AI disclosures can lead to a ``transparency dilemma'', where disclosure reduces readers' trust. However, little is known about how the \textit{level of detail} in AI disclosures influences trust and contributes to this dilemma within the news context. In this 3$\times$2$\times$2 mixed factorial study with 40 participants, we investigate how three levels of AI disclosures (none, one-line, detailed) across two types of news (politics and lifestyle) and two levels of AI involvement (low and high) affect news readers' trust. We measured trust using the News Media Trust questionnaire, along with two decision behaviors: source-checking and subscription decisions. Questionnaire responses and subscription rates showed a decline in trust only for detailed AI disclosures, whereas source-checking behavior increased for both one-line and detailed disclosures, with the effect being more pronounced for detailed disclosures. Insights from semi-structured interviews suggest that source-checking behavior was primarily driven by interest in the topic, followed by trust, whereas trust was the main factor influencing subscription decisions. Around two-thirds of participants expressed a preference for detailed disclosures, while most participants who preferred one-line indicated a need for detail-on-demand disclosure formats. Our findings show that not all AI disclosures lead to a transparency dilemma, but instead reflect a trade-off between readers' desire for more transparency and their trust in AI-assisted news content. 
\end{abstract}

\maketitle

\section{Introduction}

Since the advent of artificial intelligence (AI) tools such as ChatGPT and Midjourney, these technologies have become increasingly integrated into our daily lives, and journalism is no exception. Studies have demonstrated that these technologies can be utilized throughout the entire journalistic reporting process, from gathering to producing, verifying, and distributing news~\cite{beckett2023generating, cools2025automation}. Tasks previously requiring human judgment, such as headline optimization, lead paragraph construction, and background research synthesis, are increasingly delegated to or augmented by AI systems~\cite{d2025ai, moller2025reinforce}.

The rapid integration of AI tools in journalistic processes has led to growing calls for transparency regarding AI use in the form of AI labels and disclosures~\cite{piasecki2024ai, zier2024labeling, el2024transparent}. Currently, such AI disclosures are self-regulated and are not standardized \cite{venkatraj2025understanding}. Hence, studies have explored disclosure designs to meet readers' transparency needs and build trust. However, recent studies~\cite{morosoli2025transparency, toff2025or, longoni2022news, altay2024people, nanz2025ai} have observed a ``transparency dilemma'' with AI disclosures, where instead of increasing readers' trust, disclosures lead to lower trust. 

AI use disclosures may include details such as whether the content was AI-generated or AI-edited, as well as the steps in which AI was involved. Recent works~\cite{gamage2025labeling, chen2025examining} on labels for AI-generated content on social media demonstrated the viability of detailed disclosures in improving transparency and users' trust. However, it remains unclear how the level of detail in AI disclosures within a news-reading context affects readers' trust. Specifically, this study asks \textbf{(RQ)}: whether detailed AI disclosures mitigate or exacerbate the transparency dilemma in news.

We explore this gap through a 3 $\times$ 2 $\times$ 2 mixed factorial study involving three levels of AI disclosures (none, one-line, detailed) across two types of news (politics and lifestyle) and two levels of AI involvement (low and high). We included two types of news because previous works~\cite{morosoli2025transparency, altay2024people, nanz2025ai} suggest that this could affect readers' trust when disclosing AI use. The studies investigating the transparency dilemma in news typically focus on fully AI-generated labels. Moving away from full AI authorship, we considered two levels of AI involvement to reflect the fact that AI can be used at different stages of news production and to varying degrees~\cite{coolstransparency}. In our study, one-line disclosures indicated whether AI was used for partial content generation or final editing, while detailed disclosures further described the specific production steps involving AI, confirmed human editorial oversight, and included contact information for error reporting. We evaluated readers' trust through the News Media Trust questionnaire~\cite{stromback2020news} at both article and outlet levels, source-checking behavior, and subscription behavior. Additionally, we conducted semi-structured interviews to gain insights into the participants' perceptions of the disclosures, their preferences, and underlying reasons for their source-checking and subscription behaviors.

We found that detailed disclosures led to lower trust questionnaire scores and lower subscription rates, whereas one-line and no disclosure conditions yielded similar trust scores and subscription rates. We also found that, on average, participants checked sources the least in the no disclosure condition, followed by one-line disclosure, and the most in the detailed disclosure condition. Although AI involvement did not show considerable differences in trust questionnaire responses, high AI involvement often led to more source-checking and lower subscription, indicating that the level of AI involvement could contribute to the transparency dilemma. Results from our interview analysis showed that often interest prompted participants to check sources, whereas trust was the determining factor for subscription behavior. Interestingly, around two-thirds of the participants preferred detailed disclosures because of more transparency, and among those who preferred one-line, most of them expressed a desire for detail-on-demand disclosure designs. Our findings suggest that not all AI disclosures lead to a transparency dilemma, but detailed disclosures do. However, detailed disclosures are more aligned with the transparency expectations of the readers, highlighting a paradoxical trade-off between trust and transparency.  


\section{Related Work}

\subsection{Regulatory Responses to Generative AI and Self-regulation of Disclosure Practices}

The rapid proliferation of generative AI has prompted regulatory responses at both regional, national, and international levels across the world. Major legislative frameworks have emerged, including the EU AI Act, the US AI Bill of Rights, and UNESCO's Ethics on AI, each attempting to establish governance structures for responsible AI development and deployment~\cite{piasecki2024ai}. The AI Act, for example, proposes a framework by categorizing systems according to risk levels and imposing heightened AI disclosure obligations, including transparency, human oversight, and technical documentation~\cite{el2024transparent}. However, editorial processes and content generation typically fall outside the Act's definitional scope, leaving journalistic AI use largely unaddressed by explicit regulation. Additionally, there is emphasis that news organizations should self-regulate, leaving the regulatory responsibility with the organizations themselves~\cite{helberger2023european}.

In light of this self-regulation, the Coalition for Content Provenance and Authenticity (C2PA) was founded. It represents a voluntary, industry-led technical standard for embedding cryptographic metadata in digital content, enabling verification of origin and modification history (C2PA, 2022). Regulatory challenges specific to journalism arise from structural tensions between press freedom, source protection, proprietary systems, and auditability~\cite{ananny2018seeing, diakopoulos2017algorithmic}. Global publishers navigate fragmented compliance landscapes, where divergent regional rules create operational complexity without necessarily advancing transparency goals~\cite{quintais2025generative}.

\subsection{Audience Perspectives and AI-usage  Disclosure Dynamics in Journalism}

Scholarly and industry discourse on AI in journalism has centered predominantly on system design and producer perspectives, leaving audience mental models, literacy, and consent practices under-examined~\cite{wolker2021algorithms, thurman2019my}. Where research addresses public reception, studies on trust and acceptance reveal systematic variation by task type and perceived stakes. Audiences express greater acceptance of AI for routine, low-stakes tasks, such as weather summaries, financial updates, or sports recaps, while penalizing opaque or high-stakes applications involving investigative reporting, opinion formation, or sensitive subject matter~\cite{graefe2018readers, waddell2019can}.

Disclosure practices have the potential to mediate trust outcomes, though efficacy depends on specificity and actionability~\cite{altay2024people, morosoli2025public}. Research examining disclosure practices demonstrates that label design and information architecture critically shape audience comprehension and trust calibration. Generic labels such as ``This article was created using AI'' are frequently misinterpreted or ignored, failing to convey the nature, extent, or oversight of automated involvement~\cite{altay2024people}. Provenance cues, such as detailed bylines indicating which tasks were AI-assisted, links to methodology pages, or embedded content credentials, demonstrate improved comprehension and trust calibration when implemented in context-dependent rather than blanket form~\cite{venkatraj2025understanding}).

Recent work found that while audiences express a strong desire for transparency about AI use, they simultaneously demonstrate trust in established news organizations to use AI responsibly without requiring granular technical details~\cite{venkatraj2025understanding}). This paradox manifests in audience preferences for general organizational statements about AI policies rather than article-level disclosures for routine tasks, with trust derived primarily from institutional reputation rather than understanding of underlying technologies.

\subsection{Transparency Dilemma of AI-usage Disclosures in News}

Although transparency through AI labels and disclosures has been widely advocated, a growing body of empirical research suggests that such transparency efforts may paradoxically erode trust: a phenomenon often described as a transparency dilemma~\cite{schilke2025transparency, morosoli2025transparency}. 

This effect was first observed with news headlines labeled as AI-generated. One study~\cite{longoni2022news} found that participants consistently rated AI headlines as less accurate than human-written headlines, even when the content is factually correct, indicating a general skepticism toward AI-written news. Another study~\cite{altay2024people} found a similar effect, labeling headlines as ``AI-generated'' reduced perceived accuracy and willingness to share, regardless of veracity of the headline and actual source (human or AI). The authors attributed this AI aversion to the interpretation of AI-generated labels as content entirely produced by AI without human supervision. These studies suggest that readers have an aversion towards news content when AI usage is disclosed.

Recently, research has focused on the transparency dilemma effect beyond headlines. One study~\cite{toff2025or} using AI-generated news text content found that readers perceive articles labeled as AI-generated as less trustworthy, even though they are evaluated as accurate and fair. They also found that this effect was mitigated by disclosing the sources used to create the articles. In another study~\cite{morosoli2025transparency} involving news headlines with an image, news labeled as AI-generated were perceived as less credible, regardless of whether it was a political or non-political topic. However, the participants' willingness to share was lower only for political news labeled as AI-generated, suggesting topic sensitivity influences the effects of disclosure. These studies indicate that the negative effects of AI disclosures go beyond perceived inaccuracy and affects credibility and trustworthiness.

AI disclosures affect not just articles, but also outlets. A study investigated participants' perceptions of AI news outlets compared to trained journalists. They found that AI-generated news outlets were trusted less, especially political news outlets (compared to entertainment outlets). Moreover, participants showed lower willingness to accept advertisements from AI outlets. Their findings indicate that the transparency dilemma associated with AI disclosure extends to outlets, with potential economic implications.

\section{Methods}
In this study, we address the question: Do detailed AI disclosures mitigate or exacerbate the transparency dilemma in text-based news? We conducted a lab study following a $3 \times 2 \times 2$ mixed factorial design to investigate how disclosure levels, news type, and AI involvement influence participants' perceived trust. The study included one between-subjects factor and two within-subjects factors, as described below:

\begin{itemize}
    \item Disclosure level (between-subjects): We studied three disclosure conditions (no disclosure, one-line disclosure, and detailed disclosure), each differing in the amount of detail about AI use.
    
    \item News type (within-subjects): Participants read both political (hard news) and lifestyle (soft news) articles. Prior work~\cite{morosoli2025public, nanz2025ai} suggests that trust when disclosing AI use may be influenced by the type of news, especially political news. 
    
    \item AI involvement (within-subjects): All articles involved AI, categorized as low (minor edits to the source material) or high (substantial AI-generated content based on the source material). The process for generating articles with two levels of AI involvement is detailed in Section~\ref{sec:stimuli}. These two levels of AI involvement go beyond the fully AI-generated use case, and acknowledge that AI can be involved to varying degrees in news production~\cite{coolstransparency}.
\end{itemize}

\subsection{News Stimuli}
\label{sec:stimuli}
We selected six source news articles (three from politics and three from lifestyle) from a selection of mainstream news organizations (e.g., BBC, CNN, NOS). Each of these six articles was edited to create two versions, representing low and high AI involvement, using ChatGPT-4o.

The Low AI involvement version was generated using the prompt below, and the resulting articles were minimally edited to ensure consistency with the source material.

\begin{center}
    \begin{tabular}{p{14cm}}
{\large $\blacktriangleright$} \textit{Generate one short news article around 250-300 words based solely on the article. Make very small tweaks in the main text based on `SOURCE'. You are allowed to change the title. }
    \end{tabular}
\end{center}

The High AI involvement version was generated using the following prompt, and the generated articles were used without additional human edits, except for deleting a paragraph when necessary to maintain the target word count.

\begin{center}
    \begin{tabular}{p{14cm}}
{\large $\blacktriangleright$} \textit{Generate one short news article around 250-300 words, have a title, an introduction and specific text on `TOPIC'. Make sure it is in the form of coherent article that is based on the following URL: `SOURCE'.} 
    \end{tabular}
\end{center}

\begin{figure}
    \centering
    \includegraphics[width=0.75\linewidth]{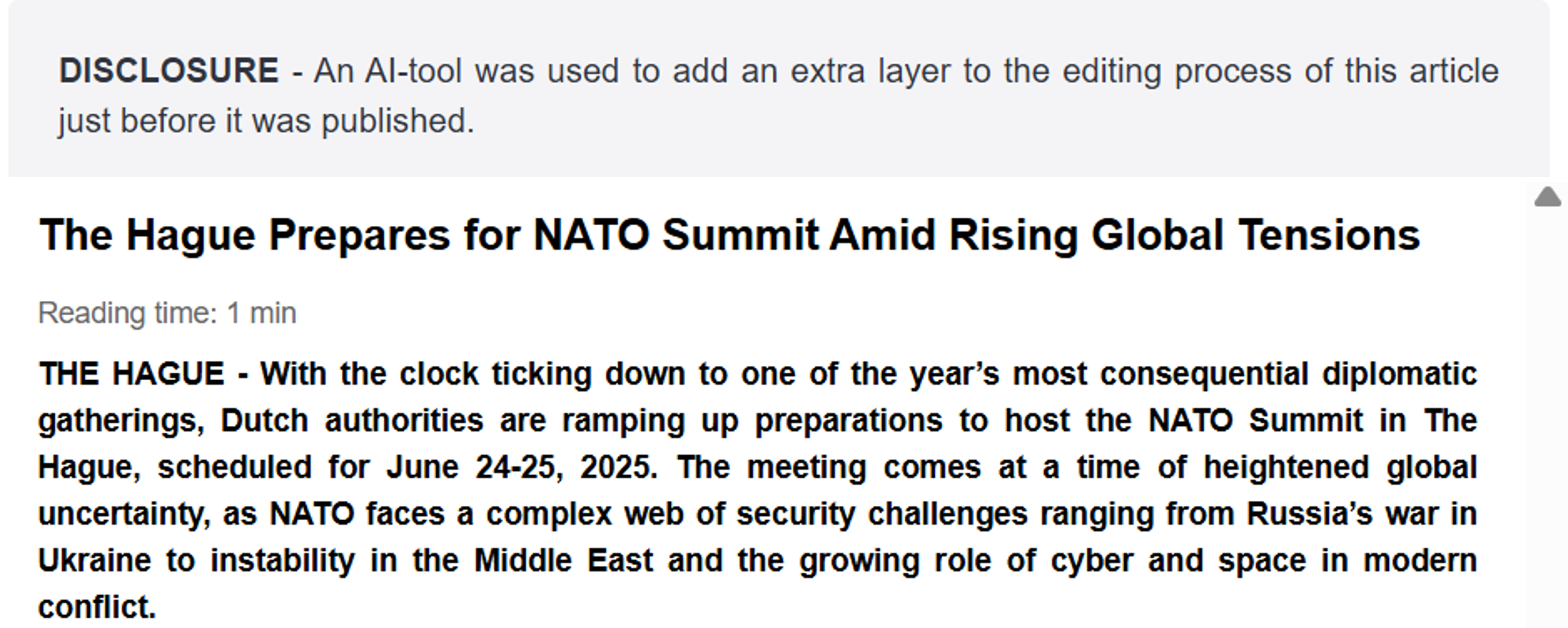}
    \caption{An example of one-line disclosure for a Low AI news article}
    \label{fig:disclosure_example}
    \Description[Example of one-line disclosure]{A one-line disclosure sits on top, informing the participant that an AI tool was used to add an extra layer to the editing process. The disclosure statement is followed by a news headline and text content.}
\end{figure}

\subsection{Levels of Disclosure}
We implemented three disclosure conditions: no disclosure, one-line disclosure, and detailed disclosure. The disclosure statements were adapted from emerging journalistic transparency practices (e.g., Mirror-UK and DR-Danish Public Broadcaster). The one-line and detailed disclosures were tailored to reflect the level of AI involvement (Low or High) in each article. The four disclosure statements (2 level of detail $\times$ 2 AI involvement) are as follows:

\begin{center}
    \begin{tabular}{p{14cm}}
    \cellcolor{gray!15} \\[-10pt]
    \cellcolor{gray!15}
    \textbf{One-line Low AI}\\
    \cellcolor{gray!15} An AI tool was used to add an extra layer to the editing process of this article just before it was published.\vspace{5pt}\\
    \end{tabular}
\end{center}

\begin{center}
    \begin{tabular}{p{14cm}}
    \cellcolor{gray!15} \\[-10pt]
    \cellcolor{gray!15}
    \textbf{One-line High AI}\\
    \cellcolor{gray!15} This article was largely generated with the help of an AI tool.\vspace{5pt}\\
    \end{tabular}
\end{center}

\begin{center}
    \begin{tabular}{p{14cm}}
    \cellcolor{gray!15} \\[-10pt]
    \cellcolor{gray!15}
    \textbf{Detailed Low AI}\\
    \cellcolor{gray!15} This article was produced with the assistance of an AI tool, which was used to support various stages of the editorial process, including content structuring, language refinement, and fact-check suggestions. All content was reviewed, edited, and approved by a human journalist before publication. This use of AI aligns with our commitment to transparency and responsible innovation in journalism. You can report errors at: webhomepage@news.com.\vspace{5pt}\\
    \end{tabular}
\end{center}

\begin{center}
    \begin{tabular}{p{14cm}}
    \cellcolor{gray!15} \\[-10pt]
    \cellcolor{gray!15}
    \textbf{Detailed High AI}\\
    \cellcolor{gray!15} This article was primarily generated using an AI tool, which was responsible for drafting substantial portions of the content, including initial story development, writing, and factual synthesis. Human editors reviewed the final version prior to publication. Our use of AI is guided by principles of transparency, accountability, and editorial oversight. Readers can report errors at: webhomepage@news.com.\vspace{5pt}\\
    \end{tabular}
\end{center}

The wording and presentation of the disclosures were pilot-tested with a small group (N=4). Following insights from previous works~\cite{longdisclosure, coolstransparency}, disclosures were displayed at the top of the article (see Figure~\ref{fig:disclosure_example}) so that participants were informed upfront about the AI’s role. Additionally, participants were instructed to read the disclosure before proceeding to the news article.

\subsection{Measures}
We collected multimodal data to capture participants' perceptions and experiences while reading the news articles. The data included self-reported questionnaires for direct subjective assessments, decision-making tasks to evaluate disclosure-related influences on decision behavior, eye-tracking data to record real-time responses, and semi-structured interviews to explore participants' reflections. Through this approach, we assess both explicit and implicit responses to AI disclosures in news content.

\subsubsection{Questionnaires}
\label{sec:questionnaires}
To study trust in this context, we used the adapted News Media Trust questionnaire~\cite{stromback2020news}. The News Media Trust questionnaire measures whether readers perceive news reporting as fair, unbiased, accurate, telling the whole story, and separating facts from opinions. Given the short length of our news stimuli, similar to previous studies~\cite{toff2025or, nanz2025ai}, we included only the fair, unbiased, and accurate sub-scales. Participants rated each sub-scale on a 5-point Likert scale ranging from Strongly disagree (1) to Strongly agree (5). As highlighted by the authors, this questionnaire can be applied at various levels, including specific media content, news organizations, or the general news media landscape of a country. In our study, the participants completed the adapted questionnaire at both the article level (immediate trust in the article just read) and the organization level, where all articles within a single session were presented as being published by the same fictional news outlet (see Figure~\ref{fig:protocol}). 


\subsubsection{Decision-making Tasks}
\label{sec:decision_tasks}
Previous studies~\cite{nanz2025ai, gamage2025labeling} suggest that informing people about AI involvement not only affects trust scores but also behaviors such as willingness to watch ads or share content. To investigate how AI disclosures influence trust-related decisions, we incorporated two decision-making tasks in our study: spending tokens to view sources and subscribing to the news outlet. 

\begin{figure}
    \centering
    \includegraphics[width=0.6\linewidth]{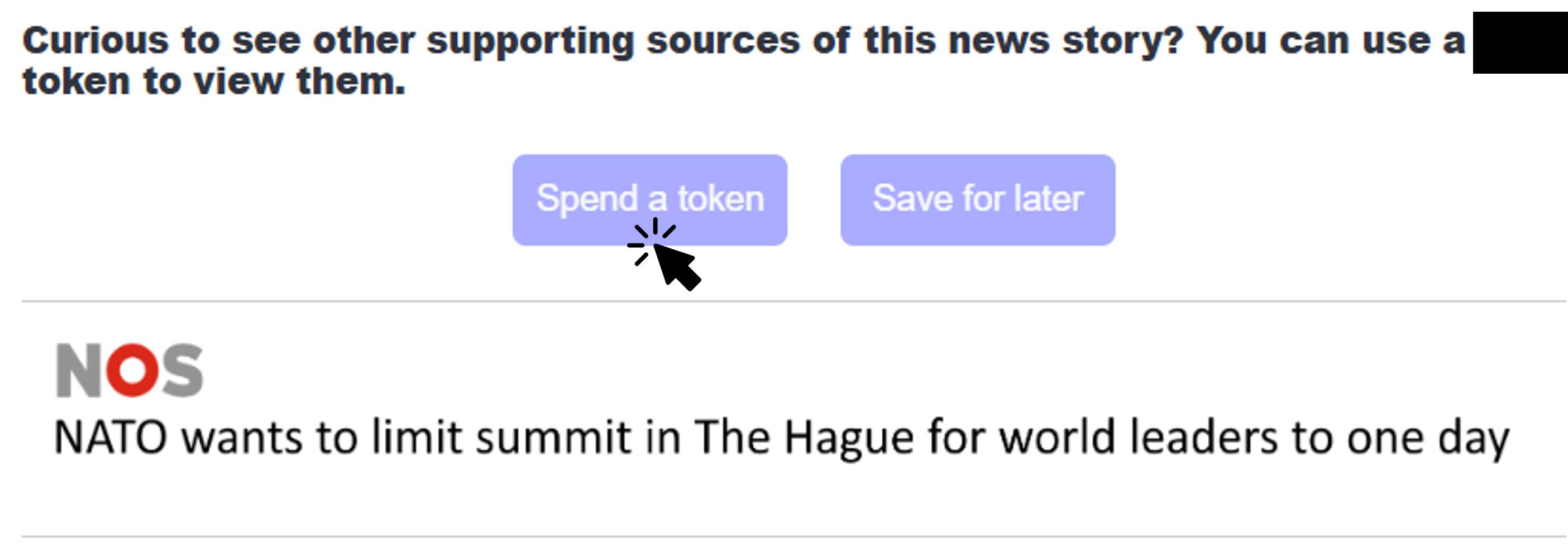}
    \caption{A screenshot of the token-spending decision task and the corresponding revealed source}
    \label{fig:reveal_source}
    \Description[Source revealed when Spend a token button is clicked]{The mouse pointer shows Spend a token button being clicked. Below, the brand logo and published source headline appear.}
\end{figure}

We used token spending as a proxy for source-checking behavior. At the beginning of the experiment, participants were informed that they had a limited number of tokens and could choose to spend a token to reveal the supporting source for any news article. This task was inspired by prior works~\cite{kim2024m, do2025hide}, which showed that participants’ tendency to seek sources serves as a proxy for lower trust. After reading each article, participants were prompted to decide whether they wanted to spend a token. If they chose to do so, they were shown the logo of the news organization and the headline of the source article, as shown in Figure~\ref{fig:reveal_source}. Although tokens were unlimited and participants could reveal sources for all articles, they were informed that tokens were limited and had to decide which articles to spend them on.

At the end of each session, participants completed a second decision-making task, in which they were asked whether they would subscribe to the news outlet that published the articles they had just read (see Figure~\ref{fig:protocol}). This binary decision was designed to study the economic implications of AI disclosures, as prior research suggests that such disclosures can influence willingness to pay and news consumption behavior~\cite{nanz2025ai, wolker2021algorithms}.



\subsubsection{Semi-structured Interview}
We conducted a short semi-structured interview (around 5 minutes) after each session to capture participants' immediate impressions about the articles and disclosures, as well as the factors influencing their choice in the decision-making tasks from Section~\ref{sec:decision_tasks}. After the last session, we held a longer interview (15-20 minutes), which focused on broader perceptions of AI in journalism and AI disclosures. During this interview, participants were shown all disclosure statements (one-line and detailed), including those not presented in their experimental sessions, and were asked for their impressions and preferences. All interviews were recorded using a Zoom H2n audio recorder and transcribed with Condens.

\subsection{Procedure}
\label{sec:procedure}
\begin{figure}
    \centering
    \includegraphics[width=0.7\linewidth]{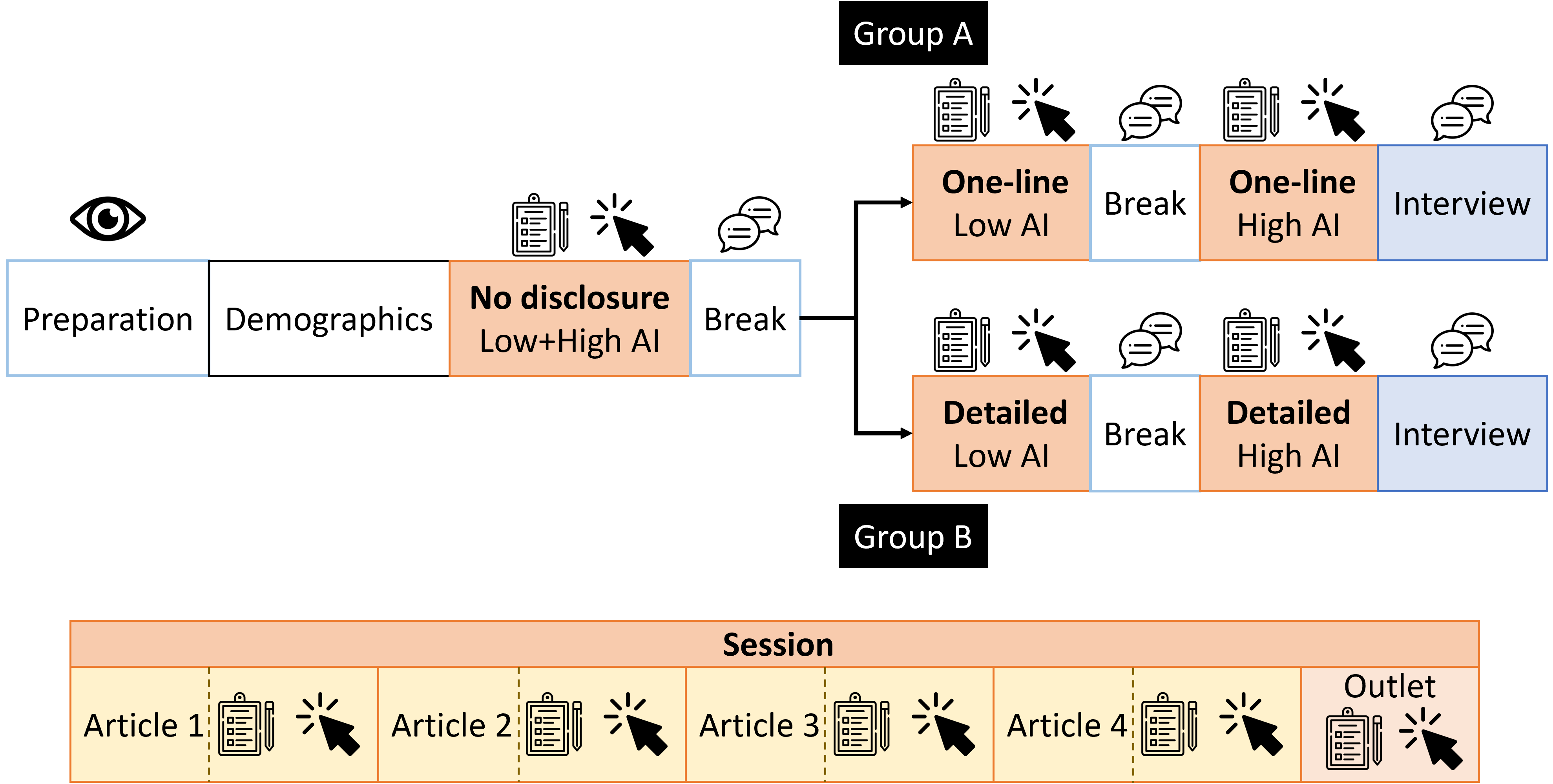}
    \caption{Overview of the experimental procedure. 
    All participants had the no disclosure condition as the first session, while the order of low- and high-AI disclosure sessions (Sessions 2 and 3) was alternated and counterbalanced across participants. Each session consisted of four articles and associated questionnaires and ended with questions about the outlet that published them.}
    \label{fig:protocol}
    \Description[Protocol for group A and B]{The initial part of the study is the same for both groups. Session 2 and 3 for group A is with one-line disclosure, and for group B is with detailed disclosure.}
\end{figure}

Our study procedure is visualized in Figure~\ref{fig:protocol}. The experiment began with a preparation phase, where participants first provided informed consent. They were then introduced to the news-reading interface and the decision-making tasks.
The participants accessed a locally hosted web interface, where they would read the news articles and complete the associated questions. On the first page, they provided demographic information, including their age range, gender, and news consumption habits. 

The study consisted of three sessions: S1, S2, and S3. In each session, participants read four news articles and completed the questionnaires about their perceptions of the article, as well as the token-spending task. Each session included two political and two lifestyle articles. The articles for each session were assigned using a Latin square rotation, and their order of presentation was randomized but alternated between lifestyle and political topics. 

Participants were randomly assigned to one of two groups: Group A or Group B. Both groups read articles with no disclosure in S1. In S2 and S3, Group A read articles with one-line disclosures, while Group B saw articles with detailed disclosures. To specifically investigate the effect of AI involvement on disclosures, all articles in S2 were either low- or high-AI versions, and the corresponding high- or low-AI versions were presented in S3. The assignment of low or high AI to S2 and S3 was alternated and counterbalanced across participants. 

Participants were told that all articles in one session were published by the same fictional news organization. At the end of each session, they completed the questionnaire on trust in the outlet and the subscription decision task. 

Each session lasted approximately 10 minutes, followed by a 5-minute break during which a short semi-structured interview was conducted. After the third session, participants took part in a longer interview (15-20 minutes). The entire experiment lasted about one hour, concluding with a debriefing session.

\subsection{Participants}
We recruited 40 participants from around the *** Institute campus through flyers, emails, and word of mouth. The sample included students, researchers, and non-scientific staff. Participants were compensated 10 USD/Euros for their time. The study was approved by the ethical committee of ***.

A power analysis using G*Power suggested that 36 participants were needed for a repeated measures within-between factors F-test to detect a medium effect size (f = 0.25) with $\alpha$ = 0.05 and a power ($1 - \beta$) of 0.95. However, during the interviews, six participants mentioned that they did not notice or read the disclosure statement for at least one session. These participants were excluded from all further analyses, resulting in a final sample of 34 participants.

\subsubsection{Demographics}
The majority of participants (n = 22, 64.7\%) were aged 25-34. Four participants (11.8\%) were in the 35-44 age range, and two participants (5.9\%) each fell into the 18-24, 45-54, 55-64, and 65 or older categories. Among the 34 participants, 21 identified as male (61.8\%), 12 as female (35.3\%), and 1 as non-binary (2.9\%).

\subsubsection{News Consumption Habits}

Most participants (n = 19, 55.9\%) reported consuming news multiple times a day, while five (14.7\%) did so once a day. Six participants (17.6\%) consumed news a few times a week, two (5.9\%) once a week, one (2.9\%) less than once a week, and one (2.9\%) rarely or never.

When asked about their primary news sources, 38.2\% (n = 13) relied on social media, followed by 35.3\% (n = 12) who used online news websites. Fewer participants got their news from video platforms (n = 3, 8.8\%), news aggregator apps (n = 2, 5.9\%), podcasts or radio (n = 1, 2.9\%), print newspapers or magazines (n = 1, 2.9\%), and other sources (n = 2, 5.9\%).

\section{Quantitative Results}
We collected three quantitative measures: News Media Trust questionnaire, source-checking behavior, and subscription behavior. We calculated the mean and standard deviation for all these measures, and ran statistical analyses using generalized linear models (GLMs) for measures showing promising differences in means. Given the exploratory nature of our study, we adjusted p-values using the Benjamini-Hochberg correction~\cite{benjamini1995controlling} to account for potential false discoveries due to multiple comparisons.

\subsection{Questionnaires}

On average, article-level news media trust scores were slightly higher for the one-line disclosure condition (mean = 10.430), followed by no disclosure (mean = 10.257) and detailed disclosure (mean = 9.938). Figure~\ref{fig:article_trust_result} shows the mean trust scores across disclosure conditions, AI involvement, and news types.

Overall, trust scores were not considerably affected by AI involvement for both political and lifestyle articles. For political articles, trust scores were relatively consistent across disclosure conditions (no disclosure: mean = 10.573; one-line: mean = 10.703; detailed: mean = 10.694). However, lifestyle articles showed a notable drop in trust scores under the detailed disclosure condition (mean = 9.180) compared to no disclosure (mean = 9.939) and one-line disclosure (mean = 10.156). These differences among lifestyle articles were statistically significant (no vs. detailed: z = 2.269, adjusted-p = 0.023; one-line vs. detailed: z = 2.827, adjusted-p = 0.040).

At the outlet level (see Table~\ref{tab:outlet_trust}), trust scores followed a similar trend, with the one-line disclosure condition rated slightly higher (mean = 10.313) than no disclosure (mean = 9.912) and detailed disclosure (mean = 9.889). 

\begin{center}
\vspace{-10pt}
    \begin{tabular}{||p{14.7cm}}
    \textbf{Takeaway:} Article-level trust questionnaire shows that one-line disclosure is on par with no disclosure condition. Detailed disclosures led to lower trust for lifestyle articles.
    \end{tabular}
\end{center}


\subsection{Decision Tasks}

\begin{figure}
    \centering
    \begin{subfigure}{\textwidth}
        \centering
        \includegraphics[width=0.7\linewidth]{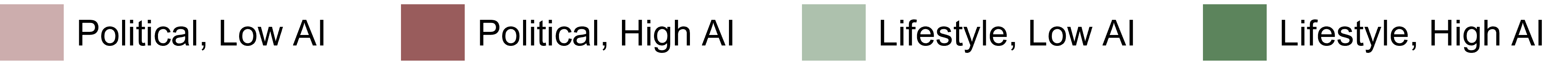}
    \end{subfigure}
    \vspace{5pt}
    \begin{subfigure}{.48\textwidth}
        \centering
        \includegraphics[width=\linewidth]{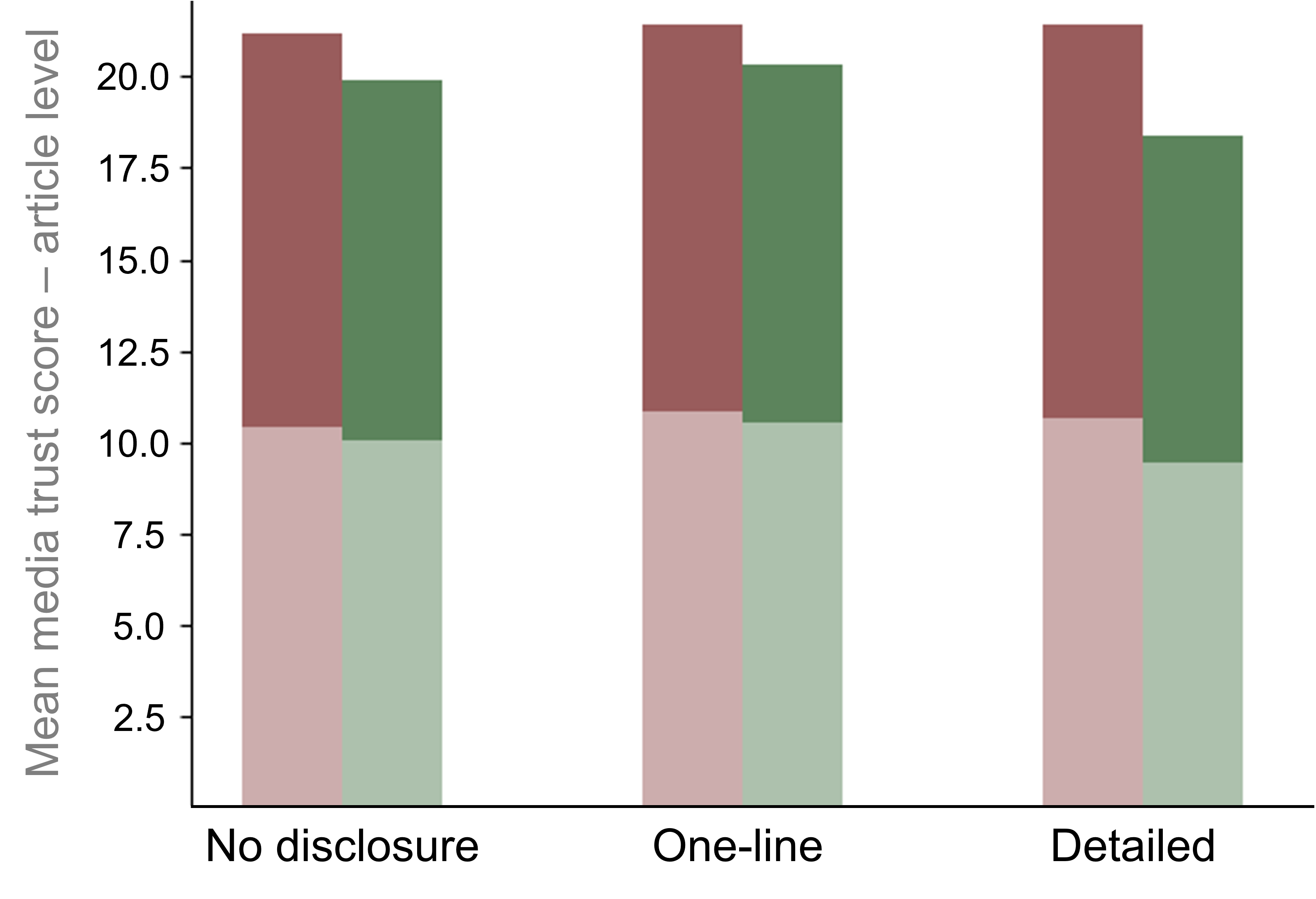}
        \caption{}
        \label{fig:article_trust_result}
    \end{subfigure}
    \hspace{5pt}
    \begin{subfigure}{.48\textwidth}
        \centering
        \includegraphics[width=\linewidth]{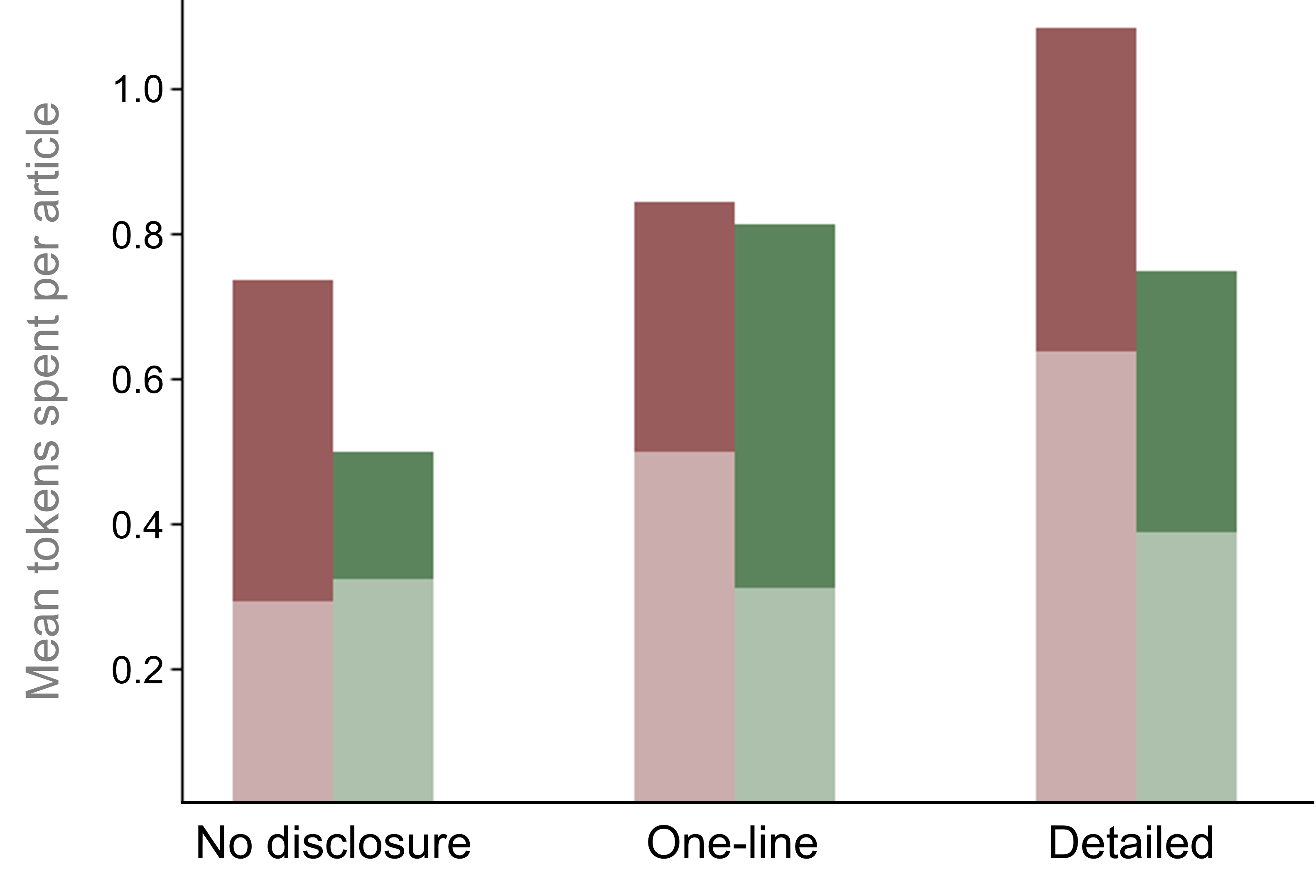}
            \caption{}
            \label{fig:token_result}
    \end{subfigure}%
    
    \caption{Stacked bar graph visualization of (a) average media trust scores (article level) across the three disclosure conditions, and (b) average token spending behavior across the three disclosure conditions.}
    \Description[Two stacked bar charts showing (a)trust scores and (b)token-spending]{Trust scores are the same across no, one-line, and detailed disclosures, except for lifestyle shows a dip in detailed. Token spending roughly increases from no to one-line to detailed.}
\end{figure}

In our study, participants spent tokens to check sources, so higher token spending indicates more instances of source-checking. Figure~\ref{fig:token_result} shows token spending behavior for articles differing in AI involvement and news type across the three disclosure conditions. Overall, participants spent fewer tokens per article in the no disclosure condition (mean = 0.309) compared to one-line (mean = 0.415) and detailed (mean = 0.459) disclosures. 

The type of news influenced token spending, with participants spending more tokens on political news (no disclosure: mean = 0.368; one-line: mean = 0.422; detailed: mean = 0.542) compared to lifestyle news (no disclosure: mean = 0.25; one-line: mean = 0.407; detailed: mean = 0.375) across all disclosure conditions. For political news, token spending in the detailed disclosure condition was significantly higher than in the no disclosure condition (z = 2.1, adjusted-p = 0.046). In lifestyle news, one-line disclosures resulted in higher token spending compared to no disclosure, with a trend toward significance (z = 1.934, adjusted-p = 0.053).

Unlike trust scores, disclosing the level of AI use resulted in differences in token spending behavior. For low-AI lifestyle articles, token spending remained similar across disclosure conditions (no disclosure: mean = 0.324; one-line: mean = 0.313; detailed: mean = 0.389). However, low-AI political articles showed significantly higher token spending in the detailed disclosure condition compared to no disclosure (no disclosure: mean = 0.294; one-line: mean = 0.5; detailed: mean = 0.639; z = 3.082, adjusted-p = 0.021). Among high-AI news articles, the difference in token-spending was more pronounced in lifestyle articles (no disclosure: mean = 0.176; one-line: mean = 0.5; detailed: mean = 0.361), with a significant difference between the one-line and no disclosure conditions (z = 2.943, adjusted-p = 0.021). However, the difference was not significant between detailed and no disclosure conditions (z = 1.789, adjusted-p = 0.08) among high-AI lifestyle news.

As shown in Table~\ref{tab:outlet_trust}, the average subscriptions were lower when AI use was disclosed. In both one-line and detailed disclosure conditions, subscriptions were lower for high-AI sessions than for low-AI sessions. The detailed disclosure condition led to the lowest subscription rate, with the overall difference compared to no disclosure trending towards significance (z = 1.955, adjusted-p = 0.059). Notably, subscriptions for detailed disclosure sessions with high-AI articles were significantly lower than in the no disclosure condition (z = 2.133, adjusted-p = 0.046).

\begin{table*}
  \caption{Mean and Standard deviation of News Media Trust questionnaire for Outlet (out of 15), Trustworthiness rating (5-point scale), and Subscription behavior across three disclosure conditions.}
  \label{tab:outlet_trust}
  \begin{tabular}{l|c|ccc|ccc}
    \toprule
    & \multicolumn{1}{c}{\large \textbf{No}} & \multicolumn{3}{c}{\large \textbf{One-line Disclosure}} & \multicolumn{3}{c}{\large \textbf{Detailed Disclosure}}\\
    \hline
    & & Low AI & High AI & Overall & Low AI & High AI & Overall \\
    Trust in Outlet & $9.912 \pm 2.19$ & $10.438 \pm 2.12$ & $10.188 \pm 1.88$ & $10.313 \pm 2.01$ & $9.833 \pm 2.24$ & $9.944 \pm 1.65$ & $9.889 \pm 1.97$\\
    Subscribe & $0.50 \pm 0.50$ & $0.50 \pm 0.50$ & $0.375 \pm 0.48$ & $0.438 \pm 0.50$ & \cellcolor{orange!25} $0.333 \pm 0.47$ & \cellcolor{orange!25} $0.222 \pm 0.42$ & \cellcolor{orange!25} $0.278 \pm 0.45$ \\
    \bottomrule
  \end{tabular}
\end{table*}

\begin{center}
\vspace{-10pt}
    \begin{tabular}{||p{14.7cm}}
    \textbf{Takeaway:} Both one-line and detailed disclosures led to higher token-spending (source-checking), with the difference more pronounced for detailed disclosure. Subscription rate under one-line disclosure is comparable to no disclosure, and lower for detailed disclosure. High AI involvement results in more pronounced differences.\\
    \end{tabular}
\end{center}

\section{Qualitative Results}
We conducted inductive thematic analysis~\cite{braun2006using} on the semi-structured interview transcripts to identify recurring patterns. Since the interview involved critical questions regarding disclosure statements, the participants who did not notice the disclosure statements were excluded. The remaining transcripts (N=34) were coded by three researchers. First, the three researchers coded the same four transcripts independently. Then, they discussed the codes and arrived at a consensus codebook. Using this codebook, the three researchers coded the remaining transcripts, 10 transcripts each. Any new codes that came up during the later coding were discussed and included in the consensus codebook. 

\subsection{Decision-making Behavior}
During the interview, the participants were asked about their decisions and the underlying factors that influenced their choices. Most participants mentioned multiple factors for their decision, especially for the source-checking task. We had codes to capture these reasons, both for positive decisions (spending a token, subscribing) and negative decisions (not spending tokens, not subscribing). In addition, we had a code 'Trust' to capture mentions of trust and sub-scales of the News Media Trust questionnaire, such as bias and fairness. 

\subsubsection{Factors influencing source-checking}
We asked the participants about their reasons for spending a token, and when they explicitly mentioned that they did not spend a token on specific articles, we followed up with questions on why they chose not to. Most participants (N=22) mentioned interest or curiosity in the topic as a factor for spending at least one of their tokens. Trust was mentioned only by 19 participants. Participants also mentioned factors like the use of AI (N=4) and well-written (N=4). Some participants also mentioned factors relating to the token system, like spending the token because it's the last article (N=8), and trying out the token system to see how it works (N=4). Interestingly, often participants had very different reasons for spending a token on one article than on the other in the same session. For example, P40 mentioned differing reasons for political and lifestyle topics: ``I think the one about [political topic] I felt was more dubious and less accurate. And the lifestyle one I felt was more factual and less opinion-based''.

When asked about the reasons to not spend a token, participants (N=13) mentioned lack of interest as a primary factor. A few participants (N=3) mentioned that the topic didn't seem important enough to be verified or checked. Participants also mentioned factors like agreeable facts (N=4), familiarity with the news (N=3), and trust (N=2) for not spending their tokens.

\begin{center}
\vspace{-10pt}
    \begin{tabular}{||p{14.7cm}}
    \textbf{Takeaway:} Overall, interest seems to play a key role in participants' decisions to both spend tokens and not spend them. Trust was only a second reason for spending tokens and rarely a factor for not spending tokens.
    \end{tabular}
\end{center}

\subsubsection{Factors influencing subscription}
Similar to source-checking, we asked participants what led them to subscribe or not subscribe to the news outlet. Trust was the most common factor (N=13) mentioned for subscribing to the outlet in at least one of the three sessions. Participants also mentioned easy to read (N=4), topics covered (N=4), and other factors. 

A majority of the participants (N=20) mentioned lack of trust (e.g., biased or one-sided reporting) as reasons for not subscribing. AI involvement and the extent of AI involvement were also mentioned as factors for not subscribing (N=12). For example, P37 mentioned, ``I don't need to subscribe to a news outlet that makes most of the stuff with AI, then I can just get the news myself.'' In most cases, both AI involvement and lack of trust were mentioned as reasons for not subscribing. A few participants (N=4) also mentioned that they disliked some specific topics (e.g., fashion) included in the session, which led them to not subscribe. Three participants mentioned that they will never subscribe to online news outlets because they prefer newspapers and non-personalized news.

\begin{center}
\vspace{-10pt}
    \begin{tabular}{||p{14.7cm}}
    \textbf{Takeaway:} Overall, trust seems to be the determining factor for subscription. Lack of trust, followed by high AI involvement, were the primary reasons for not subscribing to the outlet.
    \end{tabular}
\end{center}

\subsection{Disclosure preferences}
During the last interview session, participants were asked about the pros and cons of the disclosure statements they saw. After, they were shown both versions of the disclosures (one-line and detailed), and asked which version they would prefer and why. In many cases, this question led to discussions about other disclosure ideas and desirable characteristics. 

\subsubsection{One-line vs. Detailed}

When asked about their preferred disclosure between one-line and detailed, a majority of participants (N=22) expressed a preference for detailed, and N=11 participants preferred one-line. One participant mentioned that their preference depends on the level of AI involvement, with high AI use warranting a more detailed disclosure. When asked if there is a scenario where the non-preferred version could be used, some of the participants who preferred detailed versions mentioned that one-line disclosures could be used for ``not important news like fashion'', which points to potential topic sensitivity of AI disclosures.

The main positive characteristic mentioned for one-line disclosures was that they were short and could be quickly read. The participants mentioned that this meant these disclosures required less processing demand and were less likely to be ignored. On the negative side, they mentioned that the AI contributions were not very clear, with possibilities of different interpretations. For instance, P3 mentioned ``[AI disclosure] was a bit general, I don't know if my mother would read an article and read AI was used as a final layer, I don't know what she would understand from that. So I think it's a bit specific for our generation that knows a bit more.'' Participants also mentioned that, compared to the detailed disclosures, the one-line disclosures lacked information that they considered important, such as contact for reporting errors and assurance that a human has approved the article.

Regarding the detailed disclosures, participants mentioned that the outlets seemed very transparent about their AI usage. They also mentioned that it gives a good idea about which steps the AI was involved in. Moreover, the inclusion of contact for reporting errors was perceived as a step towards accountability. The main negative for the detailed disclosures was the length, with participants pointing out that it was too long to read and would take up too much space, especially if reading on phones. Participants (N=6) also mentioned that detailed disclosures are likely to be skipped, and some participants drew parallels with accepting cookies without reading the entire statements.

A few participants (N=3) mentioned that they need more details, like the model of AI used and ratios of how much AI was involved in each step (e.g., AI contributed to 40\% of the research and data gathering). A couple of participants (N=3) questioned whether we need AI disclosures, especially if it is used as an editing tool. A few other participants (N=2) also mentioned that for them, it is not important what is disclosed, but rather who is disclosing, stating that they would trust certain news outlets and journalists to use AI responsibly.

\begin{center}
\vspace{-10pt}
    \begin{tabular}{||p{14.7cm}}
    \textbf{Takeaway:} A majority of participants prefer detailed disclosure over one-line. Detailed is perceived as more transparent, whereas a one-line disclosure was seen as quick to read and less likely to be ignored.
    \end{tabular}
\end{center}

\subsubsection{Disclosure ideas}

When discussing their disclosure preferences, some participants mentioned alternate disclosures that they think would be more effective. Many of these disclosure designs (N=15) had a detail-on-demand characteristic, where there is a short disclosure with a possibility of getting more details by interacting with an icon or clicking a link. The main reasoning behind this is that the reader has agency on the level of disclosure and can choose to ``dig deeper for topics they care about''. Interestingly, 8 out of 11 participants who preferred one-line disclosure suggested an alternate disclosure with a detail-on-demand feature. Three participants said they would prefer more visual disclosure instead of text, and two participants mentioned highlighting key AI contributions in the detailed version for easily glancing through the statements. One participant also mentioned having short disclosures for entirely human articles, assuring the reader that AI was not involved.

\begin{center}
\vspace{-10pt}
    \begin{tabular}{||p{14.7cm}}
    \textbf{Takeaway:} Participants who preferred one-line disclosures suggested designs with a detail-on-demand feature.
    \end{tabular}
\end{center}

\subsubsection{Disclosure expectations}

Almost all participants said they believe AI is already used in journalism and are fine with AI involvement in the editing process (e.g., rephrasing, structuring). When asked if there should be government regulations on AI disclosures in the news context, a majority of the participants (N=24) said yes, although some of them (N=5) expressed concerns about its enforcement, as it is getting harder to detect AI content. 

Some participant responses suggest that AI disclosures can make the outlets be perceived as transparent, which can be a motivation for disclosure practices. Many participants (N=17) said they would have a positive perception of news outlets disclosing AI use, as such outlets are perceived as more transparent. Two participants mentioned that there is potential for misrepresenting the extent of AI involvement and ``whitewashing'' AI use, so that the outlets appear transparent and trustworthy. 

\section{Discussion}

\subsection{Key Findings and Insights Into the Transparency Dilemma in News}

We expected the transparency dilemma to manifest in both disclosure conditions, albeit at varying degrees. Interestingly, the trust questionnaire responses and subscription behaviors indicate that only detailed disclosures lead to lower trust. However, both disclosures show higher source-checking behaviors. Although both decision behaviors were integrated as measures of trust, our qualitative findings indicate that only subscription behavior was governed by trust. Curiosity or interest in the topic was the primary reason for checking sources. In light of our qualitative findings, we infer that not all disclosures lead to lower trust. 

While prior works~\cite{altay2024people, morosoli2025transparency, nanz2025ai, longoni2022news, toff2025or} studied AI disclosures for fully generated content, our disclosures reflected two levels of AI involvement rather than fully AI-generated articles. The subscription measures show slightly lower trust when disclosing high AI involvement compared to low AI involvement. Some participants also mentioned high AI involvement as a reason for not subscribing. However, the differences were not high enough to consider statistical testing, indicating our results hold regardless of AI involvement. This observation shows that high-AI involvement was not equated with full AI-generation, plausibly because our disclosures emphasized partial involvement and avoided suggesting an absence of human oversight, which \citeauthor{altay2024people} highlighted as a reason for negative perception.

Some prior works~\cite{morosoli2025transparency, nanz2025ai} suggest that political topics exacerbate the transparency dilemma than non-political topics, whereas others~\cite{toff2025or} did not find a significant difference. Surprisingly, our trust questionnaire scores indicate that lifestyle topics were affected more than political topics. However, as our qualitative findings suggest, other factors like interest or familiarity could influence this difference.

While the quantitative results suggest that a one-line disclosure is better at mitigating the transparency dilemma, qualitative results highlight readers' preference for detailed versions and detail-on-demand disclosures. Our findings position the transparency dilemma in news as a trade-off between trust and readers' transparency needs, rather than an inevitable outcome of AI use disclosures.

Overall, our findings show that detailed AI use disclosures exacerbate the transparency dilemma \textbf{(RQ)}, though not uniformly across news types and levels of AI involvement. While our study focused on news, the transparency dilemma exists in other contexts~\cite{schilke2025transparency}, and our results may extend to other high-stakes domains where AI use is disclosed.

\subsection{Practical Implications for AI Disclosures in News}
At least two practical implications for disclosures for AI use in news are formulated. First, the findings suggest that news organizations could favor concise AI-use disclosures over detailed ones. Across both trust and behavioral measures, a one-line disclosure generally performs better than a detailed disclosure, as the latter might lead to information overload. This observation indicates that transparency about AI use in news does not always require extensive explanation to be effective; in fact, overly detailed disclosures may be counterproductive~\cite{cools2025automation}. At the same time, the results have shown that transparency should not be considered as a monolith~\cite{morosoli2025public}. For example, a news topic can influence disclosure perception, and news organizations could therefore design disclosure differently in relation to such topics, which could, in turn, induce trust.  

Second, the study challenges prevailing assumptions in the literature that position disclosure as inherently trust-reducing when algorithmic or AI involvement is made salient~\cite{altay2024people}. The present findings demonstrate that disclosure effects are contingent on disclosure form and framing rather than mere presence. One-line disclosures matched no disclosure condition on trust measures while outperforming detailed disclosures, indicating that appropriate disclosure design can maintain trust while fulfilling transparency obligations. This finding addresses the ``transparency dilemma'' facing news organizations, many of which avoid disclosing AI use due to concerns about audience distrust. By demonstrating that concise disclosures do not invariably reduce trust, this study provides empirical evidence that organizations can be transparent about AI involvement without necessarily sacrificing audience trust perceptions.

\subsection{Policy Implications}

Legislators and policy experts have advocated AI disclosures as a mechanism for transparency and accountability \cite{engelmann2023algorithmic, sloane2025systematic, el2024transparent}. In an audit of 129 AI regulations, \citeauthor{sloane2025systematic} found that 27\% of transparency mandates entail disclosure requirements. While this has been central to automated decision-making systems (ADS), where transparency supports accountability and contestability \cite{engelmann2023algorithmic}, Article 50 of the EU AI Act extends AI disclosure obligations to `limited risk' systems, such as chatbots or systems generating synthetic data \cite{edwards2021eu}. Yet, questions remain about what these disclosures entail, and what their corresponding usefulness and feasibility would be, particularly in the news domain \cite{el2024transparent}. For example, \citeauthor{el2024transparent} highlight five themes of relevance for determining appropriate disclosure, including user empowerment, practicality, and industry impact. 

In the news ecosystem, AI disclosures must promote user awareness and agency without misrepresenting or undermining trust in journalism. Our findings, which expand on the ``transparency dilemma'' \cite{morosoli2025transparency,schilke2025transparency}, show that it is possible to maintain both transparency and warranted trust. 
However, such disclosures can come at the expense of simplification, which can render information less accessible~\cite{busuioc2023reclaiming}. As observed in our qualitative analysis, participants pointed out that short disclosures may leave out important information, such as details on the outlet's AI policy or a contact for error reporting. Conversely, participants noted that longer disclosures may discourage reading, leading users to treat them like cookie banners or terms of use that are often ignored. Effective disclosure, therefore, requires balancing accessibility, informativeness, and user behavior. Studies such as this, which situate disclosure within a human-computer interaction context, are essential for informing appropriate policy making.

Our study has findings pertinent to identifying the scope of regulation. For instance, the EU AI Act determines the required regulation level based on the level of risk \cite{edwards2021eu}. Similarly, in news, participants report different expectations and needs for disclosure across different domains. In both quantitative and survey results, participants indicated a greater interest in disclosure and source-checking for political news, which they deemed to have higher stakes than lifestyle news presented in the study. For higher risk domains, for example, past regulation suggests the potential of combining disclosure with other mechanisms, such as human assessment (or human-in-the-loop) \cite{sloane2025systematic}. Finally, amidst different misperceptions and heuristics about AI, disclosure is most helpful when readers are informed about what low and high modes of AI usage in journalism actually entail, and how that influences their trust and judgment of the content.

\subsection{Limitations and future work}

Several limitations of this study should be acknowledged. First, our participant pool was drawn from an academic institute, where all participants had some prior experience with AI tools. Moreover, almost all participants believed that AI is already involved to some degree in news production. As a result, the findings may not generalize to the broader population of news readers with varying levels of AI familiarity. Second, we did not examine factors such as the positioning of disclosures and the presence of established news brand logos. which could also influence readers’ trust and engagement. Third, our study focused exclusively on text-based news articles, so the effects of AI disclosures on news images or multimedia content remain unexplored.

Future research could address these limitations by investigating more diverse populations and considering additional factors. This includes exploring novel disclosure designs like detail-on-demand approaches, which could help balance transparency and trust. Furthermore, AI disclosures can be used to manipulate readers' trust, and hence, it is critical to investigate potential dark patterns in disclosure designs. Further studies involving multimedia news formats would provide a comprehensive understanding of AI disclosures and the transparency dilemma across all journalistic content.

\section{Conclusion}

Our study provides new insights into how AI disclosures in news influence readers’ trust and behavior. We show that not all AI disclosures lead to the transparency dilemma. While one-line disclosures and no disclosures yielded similar trust and subscription outcomes, detailed disclosures reduced trust and subscription rates. Disclosing AI use, regardless of the level of detail, led to an increase in source-checking behavior. Interview analysis suggests that subscription behavior was governed by trust, but source-checking was driven by curiosity and interest. Interviews further revealed that most participants either preferred the detailed disclosures or detail-on-demand disclosure designs. These findings reveal a trade-off between readers’ desire for transparency and their trust in AI-assisted journalism.

\section{Generative AI Usage Statement}
The initial draft of this paper was written by the authors without any AI assistance. Later, we used ChatGPT-5 to rephrase and shorten parts of the manuscript. We did not use any generative AI tool to create tables, figures, or perform data analysis.

\bibliographystyle{ACM-Reference-Format}
\bibliography{facct_main}

@article{schilke2025transparency,
  title={The transparency dilemma: How AI disclosure erodes trust},
  author={Schilke, Oliver and Reimann, Martin},
  journal={Organizational Behavior and Human Decision Processes},
  volume={188},
  pages={104405},
  year={2025},
  publisher={Elsevier}
}

@article{helberger2023european,
  title={The European AI act and how it matters for research into AI in media and journalism},
  author={Helberger, Natali and Diakopoulos, Nicholas},
  journal={Digital Journalism},
  volume={11},
  number={9},
  pages={1751--1760},
  year={2023},
  publisher={Taylor \& Francis}
}

@article{stromback2020news,
  title={News media trust and its impact on media use: Toward a framework for future research},
  author={Str{\"o}mb{\"a}ck, Jesper and Tsfati, Yariv and Boomgaarden, Hajo and Damstra, Alyt and Lindgren, Elina and Vliegenthart, Rens and Lindholm, Torun},
  journal={Annals of the International Communication Association},
  volume={44},
  number={2},
  pages={139--156},
  year={2020},
  publisher={Oxford University Press}
}

@article{toff2025or,
  title={“Or they could just not use it?”: The dilemma of AI disclosure for audience trust in news},
  author={Toff, Benjamin and Simon, Felix M},
  journal={The International Journal of Press/Politics},
  volume={30},
  number={4},
  pages={881--903},
  year={2025},
  publisher={SAGE Publications Sage CA: Los Angeles, CA}
}

@article{nanz2025ai,
  title={AI in the Newsroom: Does the Public Trust Automated Journalism and Will They Pay for It?},
  author={Nanz, Andreas and Binder, Alice and Matthes, J{\"o}rg},
  journal={Journalism Studies},
  pages={1--20},
  year={2025},
  publisher={Taylor \& Francis}
}

@inproceedings{gamage2025labeling,
  title={Labeling Synthetic Content: User Perceptions of Label Designs for AI-Generated Content on Social Media},
  author={Gamage, Dilrukshi and Sewwandi, Dilki and Zhang, Min and Bandara, Arosha K},
  booktitle={Proceedings of the 2025 CHI Conference on Human Factors in Computing Systems},
  pages={1--29},
  year={2025}
}

@article{wolker2021algorithms,
  title={Algorithms in the newsroom? News readers’ perceived credibility and selection of automated journalism},
  author={W{\"o}lker, Anja and Powell, Thomas E},
  journal={Journalism},
  volume={22},
  number={1},
  pages={86--103},
  year={2021},
  publisher={SAGE Publications Sage UK: London, England}
}

@article{benjamini1995controlling,
  title={Controlling the false discovery rate: a practical and powerful approach to multiple testing},
  author={Benjamini, Yoav and Hochberg, Yosef},
  journal={Journal of the Royal statistical society: series B (Methodological)},
  volume={57},
  number={1},
  pages={289--300},
  year={1995},
  publisher={Wiley Online Library}
}

@inproceedings{kim2024m,
  title={" I'm Not Sure, But...": Examining the Impact of Large Language Models' Uncertainty Expression on User Reliance and Trust},
  author={Kim, Sunnie SY and Liao, Q Vera and Vorvoreanu, Mihaela and Ballard, Stephanie and Vaughan, Jennifer Wortman},
  booktitle={Proceedings of the 2024 ACM conference on fairness, accountability, and transparency},
  pages={822--835},
  year={2024}
}

@inproceedings{do2025hide,
  title={Hide or Highlight: Understanding the Impact of Factuality Expression on User Trust},
  author={Do, Hyo Jin and Geyer, Werner},
  booktitle={Proceedings of the AAAI/ACM Conference on AI, Ethics, and Society},
  volume={8},
  number={1},
  pages={785--797},
  year={2025}
}

@article{cools2025automation,
  title={From Automation to Transformation with AI-Tools: Exploring the Professional Norms and the Perceptions of Responsible AI in a News Organization},
  author={Cools, Hannes and de Vreese, Claes H},
  journal={Digital Journalism},
  pages={1--20},
  year={2025},
  publisher={Taylor \& Francis}
}

@article{d2025ai,
  title={AI Divides in Newsrooms? How Journalists in the Low Countries Use and Perceive Generative AI},
  author={D’haeseleer, Stephanie and Van Damme, Kristin and Cools, Hannes and Van Leuven, Sarah and Evens, Tom},
  journal={Journalism Practice},
  pages={1--28},
  year={2025},
  publisher={Taylor \& Francis}
}

@article{moller2025reinforce,
  title={Reinforce, readjust, reclaim: How artificial intelligence impacts journalism’s professional claim},
  author={M{\o}ller, Lynge Asbj{\o}rn and Skovsgaard, Morten and de Vreese, Claes},
  journal={Journalism},
  volume={26},
  number={7},
  pages={1373--1390},
  year={2025},
  publisher={SAGE Publications Sage UK: London, England}
}

@inproceedings{venkatraj2025understanding,
  title={Understanding AI Disclosure Needs for News Production and Journalism},
  author={Venkatraj, Karthikeya Puttur and Morosoli, Sophie and Cools, Hannes and Naudts, Laurens and de Vreese, Claes and Helberger, Natali and Cesar, Pablo and El Ali, Abdallah},
  booktitle={Proceedings of the 24th International Conference on Mobile and Ubiquitous Multimedia},
  pages={202--208},
  year={2025}
}

@article{piasecki2024ai,
  title={AI-generated journalism: Do the transparency provisions in the AI Act give news readers what they hope for?},
  author={Piasecki, Stanislaw and Morosoli, Sophie and Helberger, Natali and Naudts, Laurens},
  journal={Internet Policy Review},
  volume={13},
  number={4},
  pages={1--28},
  year={2024},
  publisher={Berlin: Alexander von Humboldt Institute for Internet and Society}
}

@article{morosoli2025public,
  title={Public accountability and regulatory expectations for AI in journalism: qualitative evidence from focus groups with Dutch citizens},
  author={Morosoli, Sophie and Naudts, Laurens and Cools, Hannes and Venkatraj, Karthikeya and Helberger, Natali and de Vreese, Claes},
  journal={AI \& SOCIETY},
  pages={1--13},
  year={2025},
  publisher={Springer}
}

@inproceedings{longoni2022news,
  title={News from generative artificial intelligence is believed less},
  author={Longoni, Chiara and Fradkin, Andrey and Cian, Luca and Pennycook, Gordon},
  booktitle={Proceedings of the 2022 ACM Conference on Fairness, Accountability, and Transparency},
  pages={97--106},
  year={2022}
}

@article{beckett2023generating,
  title={Generating change: A global survey of what news organisations are doing with AI},
  author={Beckett, Charlie and Yaseen, Mira},
  year={2023},
  publisher={Polis, The London School of Economics and Political Science}
}

@inproceedings{zier2024labeling,
  title={Labeling AI-Generated News Content: Matching Journalist Intentions with Audience Expectations},
  author={Zier, Jessica and Diakopoulos, Nicholas},
  booktitle={Proceedings of the Computafion and Journalism Symposium 2024},
  year={2024}
}

@inproceedings{morosoli2025transparency,
  title={The Transparency Dilemma: An Experiment on How AI Disclosures Affect Credibility Perceptions and Engagement Across Topics},
  author={Morosoli, Sophie and van der Goot, Emma and Resendez, Valeria and de Vreese, Claes and Helberger, Natali},
  booktitle={Proceedings of the AAAI/ACM Conference on AI, Ethics, and Society},
  volume={8},
  number={2},
  pages={1748--1757},
  year={2025}
}

@article{altay2024people,
  title={People are skeptical of headlines labeled as AI-generated, even if true or human-made, because they assume full AI automation},
  author={Altay, Sacha and Gilardi, Fabrizio},
  journal={PNAS nexus},
  volume={3},
  number={10},
  pages={pgae403},
  year={2024},
  publisher={Oxford University Press US}
}

@inproceedings{chen2025examining,
  title={Examining the Impact of Label Detail and Content Stakes on User Perceptions of AI-Generated Images on Social Media},
  author={Chen, Jingruo and Wang, Tung-Yen and Williams, Marie and Jordan, Natalia Andrea and Shao, Mingyi and Zhang, Linda and Fussell, Susan R},
  booktitle={Companion Publication of the 2025 Conference on Computer-Supported Cooperative Work and Social Computing},
  pages={270--275},
  year={2025}
}

@article{edwards2021eu,
  title={The EU AI Act: a summary of its significance and scope},
  author={Edwards, Lilian},
  journal={Artificial Intelligence (the EU AI Act)},
  volume={1},
  pages={25},
  year={2021}
}

@article{busuioc2023reclaiming,
  title={Reclaiming transparency: contesting the logics of secrecy within the AI Act},
  author={Busuioc, Madalina and Curtin, Deirdre and Almada, Marco},
  journal={European Law Open},
  volume={2},
  number={1},
  pages={79--105},
  year={2023}
}

@inproceedings{el2024transparent,
  title={Transparent AI disclosure obligations: Who, what, when, where, why, how},
  author={El Ali, Abdallah and Venkatraj, Karthikeya Puttur and Morosoli, Sophie and Naudts, Laurens and Helberger, Natali and Cesar, Pablo},
  booktitle={Extended Abstracts of the CHI Conference on Human Factors in Computing Systems},
  pages={1--11},
  year={2024}
}

@article{sloane2025systematic,
  title={A systematic review of regulatory strategies and transparency mandates in AI regulation in Europe, the United States, and Canada},
  author={Sloane, Mona and W{\"u}llhorst, Elena},
  journal={Data \& Policy},
  volume={7},
  pages={e11},
  year={2025},
  publisher={Cambridge University Press}
}

@article{engelmann2023algorithmic,
  title={Algorithmic transparency as a fundamental right in the democratic rule of law: A comparative approach to regulation in European, North American, and Brazilian contexts},
  author={Engelmann, Alana Gabriela},
  journal={Brazilian Journal of Law, Technology and Innovation},
  volume={1},
  number={2},
  pages={169--188},
  year={2023}
}

@article{ananny2018seeing,
  title={Seeing without knowing: Limitations of the transparency ideal and its application to algorithmic accountability},
  author={Ananny, Mike and Crawford, Kate},
  journal={new media \& society},
  volume={20},
  number={3},
  pages={973--989},
  year={2018},
  publisher={SAGE Publications Sage UK: London, England}
}

@article{diakopoulos2017algorithmic,
  title={Algorithmic transparency in the news media},
  author={Diakopoulos, Nicholas and Koliska, Michael},
  journal={Digital journalism},
  volume={5},
  number={7},
  pages={809--828},
  year={2017},
  publisher={Taylor \& Francis}
}

@article{quintais2025generative,
  title={Generative AI, copyright and the AI Act},
  author={Quintais, Jo{\~a}o Pedro},
  journal={Computer Law \& Security Review},
  volume={56},
  pages={106107},
  year={2025},
  publisher={Elsevier}
}

@article{thurman2019my,
  title={My friends, editors, algorithms, and I: Examining audience attitudes to news selection},
  author={Thurman, Neil and Moeller, Judith and Helberger, Natali and Trilling, Damian},
  journal={Digital journalism},
  volume={7},
  number={4},
  pages={447--469},
  year={2019},
  publisher={Taylor \& Francis}
}

@article{graefe2018readers,
  title={Readers’ perception of computer-generated news: Credibility, expertise, and readability},
  author={Graefe, Andreas and Haim, Mario and Haarmann, Bastian and Brosius, Hans-Bernd},
  journal={Journalism},
  volume={19},
  number={5},
  pages={595--610},
  year={2018},
  publisher={SAGE Publications Sage UK: London, England}
}

@article{waddell2019can,
  title={Can an algorithm reduce the perceived bias of news? Testing the effect of machine attribution on news readers’ evaluations of bias, anthropomorphism, and credibility},
  author={Waddell, T Franklin},
  journal={Journalism \& mass communication quarterly},
  volume={96},
  number={1},
  pages={82--100},
  year={2019},
  publisher={SAGE Publications Sage CA: Los Angeles, CA}
}

@article{coolstransparency,
  title={“Transparency is More Than a Just Label": Audiences’ Information Needs for AI use Disclosures in News},
  author={Cools, Hannes and Morosoli, Sophie and Naudts, Laurens and Venkatraj, Karthikeya and de Vreese, Claes and Helberger, Natali},
  year={2025},
  publisher={OSF}
}

@article{longdisclosure,
  title={The Disclosure Dilemma: How AI Attribution Affects Reactions to Public Health Messages},
  author={Long, Jacob A and Oyewole, Tabitha and Goli, Maryam and Keisler, Jacqueline M and Alyaqout, Saud and Rodgers, Michael D and N’Diaye, Arielle}
}

@article{braun2006using,
  title={Using thematic analysis in psychology},
  author={Braun, Virginia and Clarke, Victoria},
  journal={Qualitative research in psychology},
  volume={3},
  number={2},
  pages={77--101},
  year={2006},
  publisher={Taylor \& Francis}
}

\appendix

\end{document}